\documentclass[fleqn,12pt]{elsarticle}

\usepackage{graphicx,color}
\usepackage{amsmath,amssymb}
\usepackage{slashbox}
\usepackage{feynmf}
\usepackage{young}
\newcommand{\Slash}[1]{{\ooalign{\hfil/\hfil\crcr$#1$}}}

\newcommand{\Nc}{N_{\rm c}}
\newcommand{\Nf}{N_{\rm f}}

\newcommand{\lqcd}{\Lambda_{\rm QCD}}

\newcommand{\vk}{\vec{k}}

\newcommand{\calS}{\mathcal{S}}  

\newcommand{\calA}{\mathcal{A}}

\newcommand{\calP}{\mathcal{P}}

\newcommand{\rmd}{\mathrm{d}}
\newcommand{\rmi}{\mathrm{i}}
\newcommand{\rme}{\mathrm{e}}

\newcommand{\rmsgn}{ {\rm sgn } }

\begin{document}

\begin{frontmatter}
\title{A renormalization group approach for QCD in a strong magnetic field}
\author{Toru Kojo}
\ead{torujj@physik.uni-bielefeld.de} 
\author{Nan Su}
\ead{nansu@physik.uni-bielefeld.de}
\address{Faculty of Physics, University of Bielefeld, 
D-33615 Bielefeld, Germany}
\address{{\rm (BI-TP 2013/12) } }
\begin{abstract}
A Wilsonian
renormalization group approach is applied, in order to include effects
of the higher Landau levels for quarks into a set of
renormalized parameters for the lowest Landau level (LLL),
plus a set of operators made of the LLL fields.
Most of the calculations can be done 
in a model-independent way
with perturbation theory
for hard gluons,
thanks to {\it form factors} of quark-gluon vertices 
that arise from the Ritus bases for quark fields.
As a part of such renormalization program, 
we compute the renormalized quark self-energy at
1-loop, including effects from all higher {\it orbital} levels.
The result indicates that the higher orbital levels
cease to strongly affect the LLL at a rather small magnetic field of 
$(0.1 - 0.3)\, {\rm GeV^2}$.
\end{abstract}
\end{frontmatter}

\section{Introduction}

A uniform magnetic field $B$ quantizes
quark dynamics in directions transverse to $B$,
leading to discretized Landau levels.
The quarks do not explicitly depend on
transverse momenta,
and may be regarded as quasi-two dimensional.
An energy splitting between levels is $\sim \sqrt{2|e B|}$,
and each level has a degeneracy proportional to $|eB|/2\pi$.
In particular, the degeneracy in the lowest Landau level (LLL)
allows more quarks to stay at low energy for larger $B$,
so more effects on non-perturbative 
dynamics \cite{Shovkovy:2012zn}.
They affect chiral symmetry breaking \cite{Suganuma:1990nn},
gluon polarization (or sea quark effects) \cite{Bruckmann:2013oba}, 
and also meson structures \cite{Hidaka:2012mz}
and their dynamics \cite{Buividovich:2010tn,Fukushima:2012xw}.
Quenched \cite{Buividovich:2008wf}
as well as full lattice calculations \cite{Bali:2011qj}
show that the chiral condensate depends linearly on $|eB|$
at $|eB| \gg \lqcd^2 \sim 0.04\, {\rm GeV}^2$.
This likely indicates the mass gap of a quark in the LLL
to be nearly $B$-independent at large $B$ \cite{Kojo:2012js},
unlike the QED case with the electron mass gap
of $\sim |eB|^{1/2}$.
This linear rising behavior and deviations from
chiral effective model predictions
start to occur at $|eB| \sim 0.3\, {\rm GeV}^2$
\cite{Bali:2012zg},
beyond which non-perturbative methods at microscopic level
seem to be required.

The Landau level quantization 
is only an approximate concept,
because gluons couple to different Landau levels.
At small $|eB| \ll \lqcd^2$, 
the different Landau levels are close in energies,
and are easily mixed by 
soft gluons accompanying large $\alpha_s$.
In this situation, we should instead use
effective theories for hadrons,
regarding $|eB|/\lqcd^2$ as a small expansion 
parameter \cite{Ioffe:1983ju}.

On the other hand,
the Landau levels are widely separated
at large $|eB| \gg \lqcd^2$,
and then discussions based on 
the Landau levels are supposed to be the useful starting point.
There is an expansion parameter,
$\lqcd^2/|eB|$,
and because each of the Landau levels has a characteristic
spatial wavefunction in transverse directions,
there naturally arise $B$-dependent {\it form factor effects},
with which soft gluons decouple from the hopping process 
of a quark
from one to another {\it orbital} level.
This feature can be used to set up 
a framework,
designed to analyze perturbative effects
separately from non-perturbative ones.

Our ultimate goal will be to understand
non-perturbative phenomena at strong $B$,
and for this purpose
it is most important to study quarks in the LLL
by applying some non-perturbative methods.
But for such analyses, 
we need to prepare renormalized parameters
which concisely summarize 
effects from all higher Landau levels (hLLs),
plus effective operators made of the LLL fields.
They are generated from diagrams having the LLL
fields as external legs 
(we will impose some particular gauge 
fixing\footnote{In principle
a gauge invariant effective action
for the LLL and soft gluons
can be derived by integrating out
hard gluons subject to the background gauge fixing condition.
But it makes the analyses far more complicated
so we will not attempt it.
}
that will be also used in non-perturbative analyses
for consistent treatments).
Below we present the first step/part of such a program.

Most of the computations can be done within 
perturbation theory,
thanks to the aforementioned form factor effects.
The perturbative framework is convenient 
to write down the renormalized 
parameters at finite $B$,
because we can easily relate them to the $B=0$ ones
for which many perturbative results are 
available \cite{Beringer:1900zz}. 
In this way all the ultraviolet (UV) divergences
can be handled.
Meanwhile, a coupling within the same
{\it orbital} levels is exceptional
and is beyond the scope of the perturbative framework:
The LLL and the first Landau level with
spin anti-parallel to that of the LLL
can couple via soft gluons,
because these Landau levels actually belong to the
same orbital level, $l=0$.
But even in that case, a semi-perturbative framework
is applicable:
Because a quark in the first Landau level
is hard, operator product expansion (OPE)
can be used to deal with soft gluons \cite{Shifman:1978bx},
in terms of local matrix elements of gluonic
operators with perturbative coefficients of ${\cal O}(|eB|^{-n})$.
The results for the OPE will be presented elsewhere.

As a first step of our program,
we shall compute 1-loop perturbative contributions
from all higher {\it orbital} levels to
the renormalized quark self-energy of the LLL,
that will be the input for non-perturbative analyses.
In the renormalization procedure,
it is essential to separate the LLL intermediate
states to avoid large logarithms associated with
the dimensional reduction, 
$\sim \ln^n (|eB|/m^2)$ \cite{Gusynin:1998nh}.
Unlike the $B=0$ case,
such an infrared (IR) enhancement
cannot be removed by simply imposing
the renormalization condition at
large external momenta.
This is part of the reason to use 
a renormalization group (RG) approach of the
Wilsonian type in which we integrate out only hard modes.
Our results should be also insensitive to the magnetic
catalysis whose main actors are quarks in 
the LLL\footnote{In QED, it has been argued
that magnetic catalysis, mainly driven by
electrons in the LLL, can generate a non-perturbative
Zeeman effect and energy splittings in the hLLs \cite{Ferrer:2008dy}.
In our approach,
these interesting radiative corrections
are separated together with the LLL.}.

Our usage of the RG differs 
from the previous RG studies in its motivation.
In the previous studies,
the RG methods were used to investigate 
the non-perturbative aspects
of QED or the 4-Fermi interaction models for QCD 
\cite{Fukushima:2012xw,RG},
which were intended to capture some {\it universal} aspects
of fermions in strong magnetic fields.
There the RG was mainly applied to non-perturbative LLL 
dynamics of such models.
In contrast, our present study is {\it not} for 
non-perturbative aspects:
considering difficulties in the first principle treatments
of non-perturbative gluon dynamics,
we will {\it not} apply the RG to the LLL dynamics
to which the non-perturbative gluons couple in a vital way.
We shall only investigate
when and how the LLL dominance over
the hLLs can be a good description,
by applying the RG only to the hLLs and hard gluons.

The virtue of our approach is that,
as far as $|eB|$ is large enough,
the method of RG can be combined with 
systematic (semi-)perturbative treatments of QCD
in a model-independent way.
Even within perturbative treatments,
a number of interesting conclusions can be derived.
In particular, it is possible to examine
the range of $|eB|$
where the perturbation theory is safely applied, 
and, within such a domain,
one can argue when higher orbital contributions are negligible.
Remarkable outputs in this work is that
the higher orbital levels cease to strongly affect the LLL 
already at a rather small value, 
$|eB| \simeq (0.1-0.3)\, {\rm GeV^2}$,
due to small $\alpha_s$
appearing in the functional integration of the higher orbital levels.
Beyond such value of $|eB|$,
the lowest orbital level approximation for
non-perturbative phenomena seems to 
be well justified.

\section{Ritus bases and the quark-gluon vertex}

All of our framework relies on the Ritus base,
so let us examine its Feynman rules and their
accompanying form factors.
We start with the quark part of the Euclidean action
($f$: flavor index, $e_f$: electric charge, 
$g_{\mu \nu} = \delta_{\mu \nu}$,
$\gamma_\mu^\dag = \gamma_\mu$: metric convention),
\begin{equation}
\calS_0 = \int \rmd^4 x ~ 
\bar{\psi}_f \left[ \Slash{\partial} 
+ \rmi e_f \Slash{\calA_\alpha} + m_f \right] \psi_f \,,
~~~
\calS_{ {\rm int} } 
= \int \rmd^4 x ~\bar{\psi}_f \, \rmi t_a \Slash{A}^a \psi_f 
\,,
\end{equation}
where $\calA$ and $A^a$ are $U(1)_{em}$ 
and $SU(\Nc)$ gauge fields, respectively,
and the $SU(\Nc)$ gauge coupling constant is 
included in the definition of $A^a$.
We apply an external, uniform magnetic field in $x_3$-direction,
which is given by a vector potential
$\calA_\mu = (\calA_1, \cdots, \calA_4) 
=(0, Bx_1, 0, 0)$.
Then the action $\calS_0$ can be diagonalized
using the Ritus bases.
Let us define projection operators,
\begin{equation}
\calP_\pm^{B_f}
= \frac{\, 1 \pm s_{B_f} \rmi \gamma_1 \gamma_2 \,}{2} \,,
~~~ \psi_{\pm f} = \calP^{B_f}_\pm \psi_f \,,
~~~ \rmi \gamma_1 \gamma_2 \, \psi_{\pm  f}
= \pm \, s_{B_f}  \psi_{\pm f} \,,
\end{equation}
where we introduce short handed notation,
$B_f \equiv e_f B$ and $s_{B_f} \equiv \rmsgn( B_f )$.
Below we omit flavor index $f$ until it
becomes important.
Now we expand a quark field by the Ritus bases,
\begin{equation}
\psi_\pm (x)
= \sum_{l=0} \int_{p_L, p_2}
\psi_{\pm} ( l, p_2, p_L ) \,
H_l \!\left( x_1-  s_B p_2/|B| \right) 
\,  \rme^{-\rmi p_2 x_2} \, \rme^{-\rmi p_L x_L } 
\,,
~~~ \int_{p_L, p_2} \equiv \int \frac{\rmd^2 p_L \rmd p_2 }{(2\pi)^3} \,,
\end{equation}
where $H_l$ is the harmonic oscillator base with
$m\omega = |B|$.
The index $l$ characterizes orbital levels.
Here the index $p_L$ stands for $(p_3,p_4)$,
and $p_\perp$ will be used for $(p_1,p_2)$.
Since the action is not diagonal in $l$,
we relabel quark fields as
$\chi_{p_2} (p_L) = \psi^+_{0, p_2} (p_L) 
\equiv \psi_+(l=0, p_2, p_L)$,
$\psi^+_{n, p_2} (p_L) \equiv \psi_+(l=n, p_2, p_L)$,
and $\psi^-_{n, p_2} (p_L) \equiv \psi_- (l=n-1, p_2, p_L)$.
Then we arrive at
\begin{equation}
\calS_0 
=  \int_{p_L, p_2} \left[\,
\bar{\chi}_{p_2} (p_L) \left( -\rmi \Slash{p}_L + m \right) \chi_{p_2} (p_L)
+ 
\sum_{n=1} 
\bar{\psi}_{n,p_2} (p_L) \left( -\rmi \Slash{P}_n + m \right) \psi_{n,p_2}
(p_L) \, \right]
\,,
\end{equation}
where $\psi_n \equiv \psi_{n+} + \psi_{n-}$
and $(P_n)_\mu \equiv (0, - s_B \sqrt{2 n |B|}, p_L)$.
This expression means that
the propagator is diagonal in $n$ and $p_2$,
and for each index the only variable is $p_L$. 
In this sense the quark dynamics is dimensionally 
reduced from four to two dimensions.

By adding the quark-gluon coupling,
$p_2$ is still a good quantum number,
but the Landau level index $n$ is not.
Again expanding quark fields in the Ritus bases,
a gluon field is convoluted with
a form factor function,
\begin{equation}
\int_{x_1} H_l(x_1-r_{p_2}) H_{l'} (x_1-r_{p_2-k_2}) \, 
\rme^{-\rmi k_1 x_1} 
\equiv 
\rme^{\rmi \Phi_{p_2,p_2-k_2} (k_1)  }
I_{l,l'} (\vk_\perp) \,,
~~~~ \Phi_{p_2,p_2-k_2} (k_1) = - \frac{k_1 (2p_2-k_2) }{ 2B } \,.
\end{equation}
The form factor naturally arises because each orbital level
has a characteristic spatial wavefunction in transverse directions.
It decides which gluons, soft or hard, are relevant for
each process.
The phase factor has a property
$\Phi_{p_2,p_2-k_2} (k_1) = - \Phi_{p_2-k_2,p_2} (-k_1)$
under exchanges $p_2 \leftrightarrow p_2-k_2$ and
$k_1 \leftrightarrow -k_1$,
which will be frequently used in the computation.
 
There are four types of vertices:
The first two conserve spins
and couple only to $A_L^a$,
\begin{equation}
\hspace{-0.7cm}
\calS^{++}_{ {\rm int} }
= \rmi \sum_{n,n'=0} 
\int_{p_L,p_2,k_L,k_2} \!\!
\bar{\psi}_{n,p_2}^+ (p_L ) \, \gamma_L t_a\,
\psi_{n', p_2-k_2}^+ ( p_L - k_L ) \,
\int_{k_1} A_{L}^a (k) \,
I_{n,n'} (\vk_\perp)  \,
\rme^{\rmi \Phi_{p_2,p_2-k_2} (k_1)  } \,,
\end{equation}
\begin{equation}
\hspace{-0.7cm}
\calS^{--}_{ {\rm int} }
= \rmi \sum_{n,n'=1} 
\int_{p_L,p_2,k_L,k_2} \!\!
\bar{\psi}_{n,p_2}^- (p_L ) \, \gamma_L t_a\,
\psi_{n', p_2-k_2}^- ( p_L - k_L ) \,
\int_{k_1} A_{L}^a (k) \, I_{n-1,n'-1} (\vk_\perp) \, 
\rme^{\rmi \Phi_{p_2,p_2-k_2} (k_1)  } \, ;
\end{equation}
and the others flip spins and couple only to $A_\perp^a$,
\begin{equation}
\hspace{-0.7cm}
\calS^{+-}_{ {\rm int} }
= \rmi \sum_{n=0,n'=1} 
\int_{p_L,p_2,k_L,k_2} \!\!
\bar{\psi}_{n,p_2}^+ (p_L ) \, \gamma_\perp t_a\,
\psi_{n', p_2-k_2}^- ( p_L - k_L ) \,
\int_{k_1} A_{\perp}^a (k) \, I_{n,n'-1}(\vk_\perp) \,
\rme^{\rmi \Phi_{p_2,p_2-k_2} (k_1)  } \,,
\end{equation}
\begin{equation}
\hspace{-0.7cm}
\calS^{-+}_{ {\rm int} }
= \rmi \sum_{n=1,n'=0} 
\int_{p_L,p_2,k_L,k_2} \!\!
\bar{\psi}_{n,p_2}^- (p_L ) \, \gamma_\perp t_a\,
\psi_{n', p_2-k_2}^+ ( p_L - k_L ) \,
\int_{k_1} A_{\perp}^a (k) \,
I_{n-1,n'}(\vk_\perp) \,
\rme^{\rmi \Phi_{p_2,p_2-k_2} (k_1)  } \,.
\end{equation}
Note that the spin flipping vertices
couple fields with $+$ and $-$ components.
In particular, $A_\perp^a$ kicks out
fields in the LLL to the hLLs,
so processes involving $A_\perp^a$
are suppressed by $1/|B|$.
Thus only $\chi$ and $A_L^a$ become relevant 
at low energy.

For later convenience,
we discuss properties of $I_{l,l'}(\vk_\perp)$.
Its general form is given by
\begin{equation}
I_{l,l'} (\vk_\perp)
= \exp\left( {-\frac{\, |z|^2 \,}{2} } \right)\times
\left\{ 
\begin{matrix}
&\sqrt{ \frac{\, l\,! \,}{\, l'! \,} } \, (z^*)^{l'-l} \, L_l^{(l')} (|z|^2) \,,
~~~~~~({\rm for}~l\le l') \\
&\sqrt{ \frac{\, l'! \,}{\, l\,!\,} } \, (-z)^{l-l'} \, L_{l'}^{(l)} (|z|^2) \,,
~~~~~~({\rm for}~l\ge l') \,,
\end{matrix}
\right.
\end{equation}
where $L_l^{l'}$ is the generalized 
Laguerre polynomial,
and the complex variables are
\begin{equation}
z(\vk_\perp) \equiv \frac{\, \rmi k_1 + s_B k_2 \,}{ \sqrt{2|B|} } \,,
~~~~
z^* (\vk_\perp) \equiv \frac{\, -\rmi k_1 + s_B k_2 \,}{ \sqrt{2|B|} }
\,,
~~~~ z(\vk_\perp) = - z(-\vk_\perp) \,.
\end{equation}
We are interested in 
the coupling between the LLL and hLLs,
and we frequently use the relation
\begin{equation}
I_{0,l} ( \vk_\perp ) 
= \rme^ {-\frac{\, |z|^2 \,}{2} } \,
\frac{(z^*)^l}{\, \sqrt{l!} \,}  \,,
~~~~
I_{0,l} ( \vk_\perp ) I_{l,0} ( -\vk_\perp ) 
= \rme^ {- |z|^2 \, } 
\frac{\, |z|^{2l} \,}{\, l! \,}  \,,
\label{formula}
\end{equation}
with $|z|^2 = k_\perp^2/2|B|$.

\begin{figure}[tb]
\vspace{0.0cm}
\begin{center}
  \includegraphics[scale=.25]{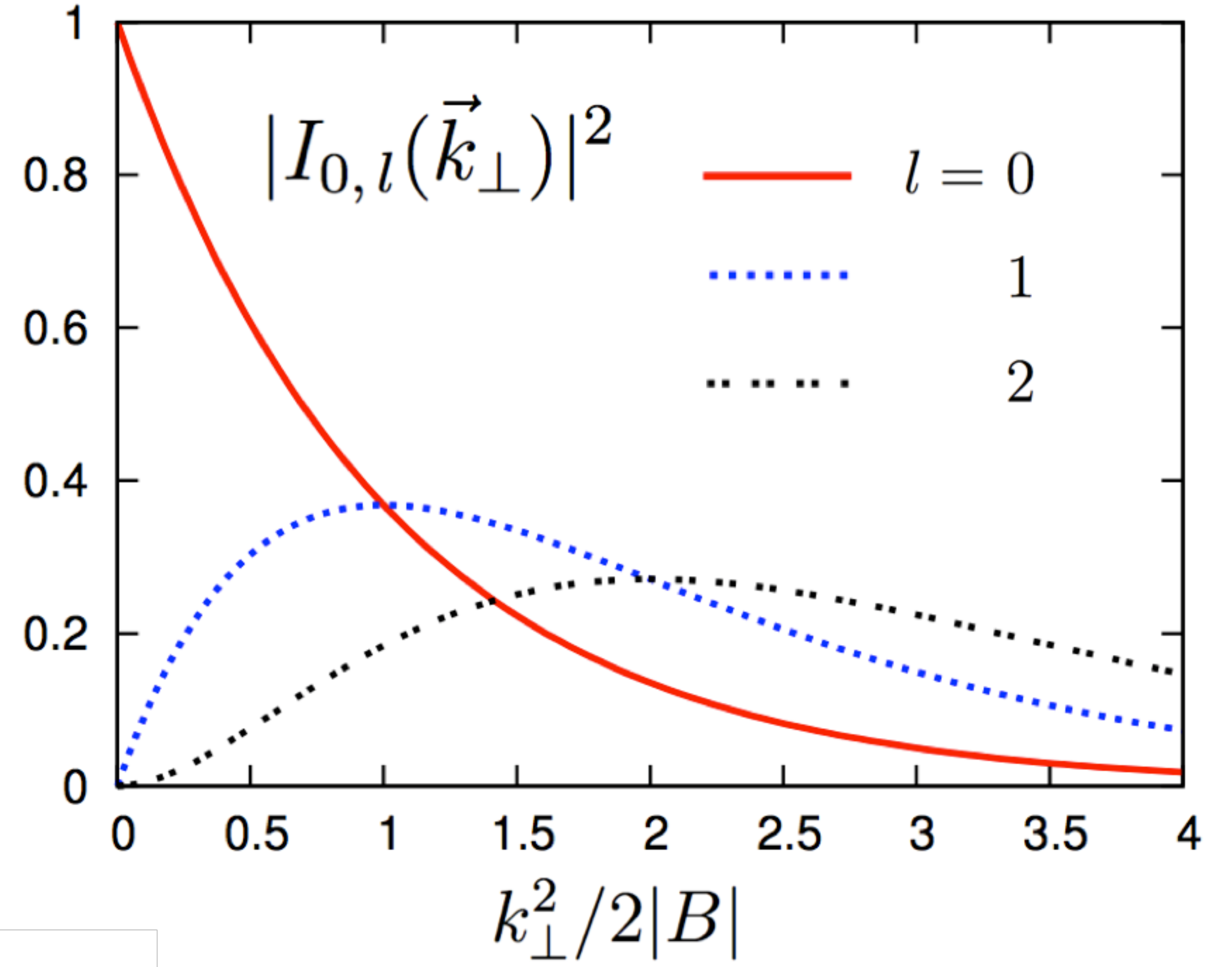} 
\end{center}
\vspace{-0.6cm}
\caption{An example of the form factor function.
We plot the $|I_{0, l}(\vk_\perp)|^2$
which will be used in the quark self-energy calculations.
There the form factor is convoluted with the gluon propagator.
The location of the maximum is $k_\perp^2 = 2l |B|$.
For $l \ge 1$, the form factor suppresses the soft gluons,
and this effect becomes stronger at larger $|B|$.
}
\label{fig:formfactor}
\end{figure}

These expressions suggest that,
for a hopping between different orbital levels,
$l$ and $l+\Delta l$,
we have to convolute the form factor function
accompanying powers of transverse momenta,
$(z/2|B|)^{ |\Delta l |/2}$
or $(z^*/2|B|)^{ |\Delta l |/2}$,
which dies out for small $k_\perp$
(see Fig.\ref{fig:formfactor}).
These powers suppress the IR contributions
of the gluon propagator,
so for $\Delta l \neq 0$
we can avoid soft gluons
and apply weak coupling methods
for hard gluons.
This observation is the base 
for our perturbative framework.

%

\section{The renormalized quark self-energy at 1-loop}

\subsection{Renormalization}

Now we use the Feynman rules in the Ritus bases
to compute the higher orbital level contributions
to the LLL quark self-energy.
To deal with the UV divergence arising from
summation of all higher orbital levels,
we use the renormalized perturbation theory
with counter terms which are determined in the $B=0$ case.
Since the LLL kinetic term does not have
$\gamma_\perp$,
we will use the self-energy expression at $B=0$ 
for $p_\perp=0$
($\Sigma^R$ and $\Sigma$ are the renormalized and
bare self-energies, respectively),
\begin{equation}
\Sigma^R_{ {\rm vac} } (p_L;\mu)
= \Sigma_{ {\rm vac} } (p_L)
- \rmi \, \delta_Z(\mu) \, \Slash{p}_L + \delta m(\mu) \,.
\label{Rvac}
\end{equation}
where $\delta_Z(\mu)$ and $\delta m(\mu)$
are counter terms at scale $\mu$
to force the renormalized self-energy to vanish at
$p_L^2=\mu^2$ (renormalization condition).
It is implicit that $\Sigma_{ {\rm vac} }$ 
depends on $\mu$ only through renormalized parameters,
$m(\mu)$, $g(\mu)$, etc.
For the self-energy at finite $B$,
we must use the same counter terms as the $B=0$ case.
Replacing the counter terms by 
$\Sigma^R_{ {\rm vac} } (p_L;\mu) - \Sigma_{ {\rm vac}
} (p_L)$ 
through (\ref{Rvac}), we have
the renormalized quark self-energy at finite $B$:
\begin{equation}
\Sigma_{B}^R (p_L; \mu)
= 
\Sigma_{ {\rm vac} }^{R} (p_L; \mu)
+ \left[\,\Sigma_{B} (p_L) 
  - \Sigma_{ {\rm vac} } (p_L) \, \right] \,.
\label{RB}
\end{equation}
The overall of the terms in the bracket is UV finite,
and vanishes as $|B| \rightarrow 0$.
The UV divergence is handled only through
the first term.

Our computation requires an additional step,
due to the need of separating the $l=0$ contribution.
To do so, we first split the bare self-energies
into two parts, depending on whether
they contain contributions
responsible for the $l=0$ level or not:
\begin{equation}
 \Sigma_{B } 
= \delta \Sigma_{B} 
+ \Sigma_{B}^{ l=0  } \,,
~~~~
 \Sigma_{ {\rm vac} } 
= \delta \Sigma_{ {\rm vac} } 
+ \Sigma_{ {\rm vac} }^{ l=0 } \,,
\end{equation}
with which we reorganize (\ref{RB}) as
\begin{equation}
\Sigma_{B}^R 
= 
\big[\, \Sigma_{ {\rm vac} }^{R} 
- \Sigma_{ {\rm vac} }^{ l=0  } \, \big]
+ \big[\, \delta \Sigma_{B}  
  - \delta \Sigma_{ {\rm vac} } \, \big] 
+ \Sigma_{B}^{ l=0 }   \,,
\label{reno1}
\end{equation}
where the first two brackets contain contributions
responsible for $l\ge 1$ levels,
while the last term is responsible for the $l=0$ level.
(How to identify $\Sigma^{l=0}_{ {\rm vac} }$
will be shown later with an explicit example in Eq.(\ref{natural}))
Therefore it is natural to {\it define}
\begin{equation}
\Sigma_{B}^{R,\,  l=0}  
\equiv \Sigma_{B}^{ l=0 }  \,,
\end{equation}
which is responsible for the $l=0$ contribution
that is left for non-perturbative studies,
and
\begin{equation}
\delta \Sigma_{B}^R 
\equiv
\delta \Sigma_{ {\rm vac} }^{R} 
+ \left[\, \delta \Sigma_{B}  
  - \delta \Sigma_{ {\rm vac} } \, \right] \,,
~~~~~
\delta \Sigma_{ {\rm vac} }^{R} 
\equiv 
\Sigma_{ {\rm vac} }^{R} 
- \Sigma_{ {\rm vac} }^{ l=0  }  \,,
\end{equation}
which include $l\ge 1$ contributions.
Note that all terms in $\delta \Sigma_{B}^R$
are expressed in terms of $\delta \Sigma$,
for which we can avoid the IR component of the gluon propagator. 
They will be the targets of our perturbative computations.

Here one might wonder why we also split the vacuum piece
into $l=0$ and the others.
Indeed, this is not a mandatory step.
However, by adjusting the phase space
for each of $B=0$ and $B\neq 0$ terms explicitly,
trivial $B$-dependence coming from phase space mismatch
can be avoided for each term, and we can extract out
genuine $B$-dependent contributions for each phase space.
This helps qualitative interpretations.
Also,
it becomes easier to see which renormalization scale
should be taken to avoid the logarithmic enhancement
typical in perturbation theory
(see Eq.(\ref{deltaSigma})).

\subsection{An example of computation at finite $B$}

%
\begin{figure}[tb]
\vspace{0.0cm}
\begin{center}
\scalebox{0.6}[0.6] {
  \includegraphics[scale=.35]{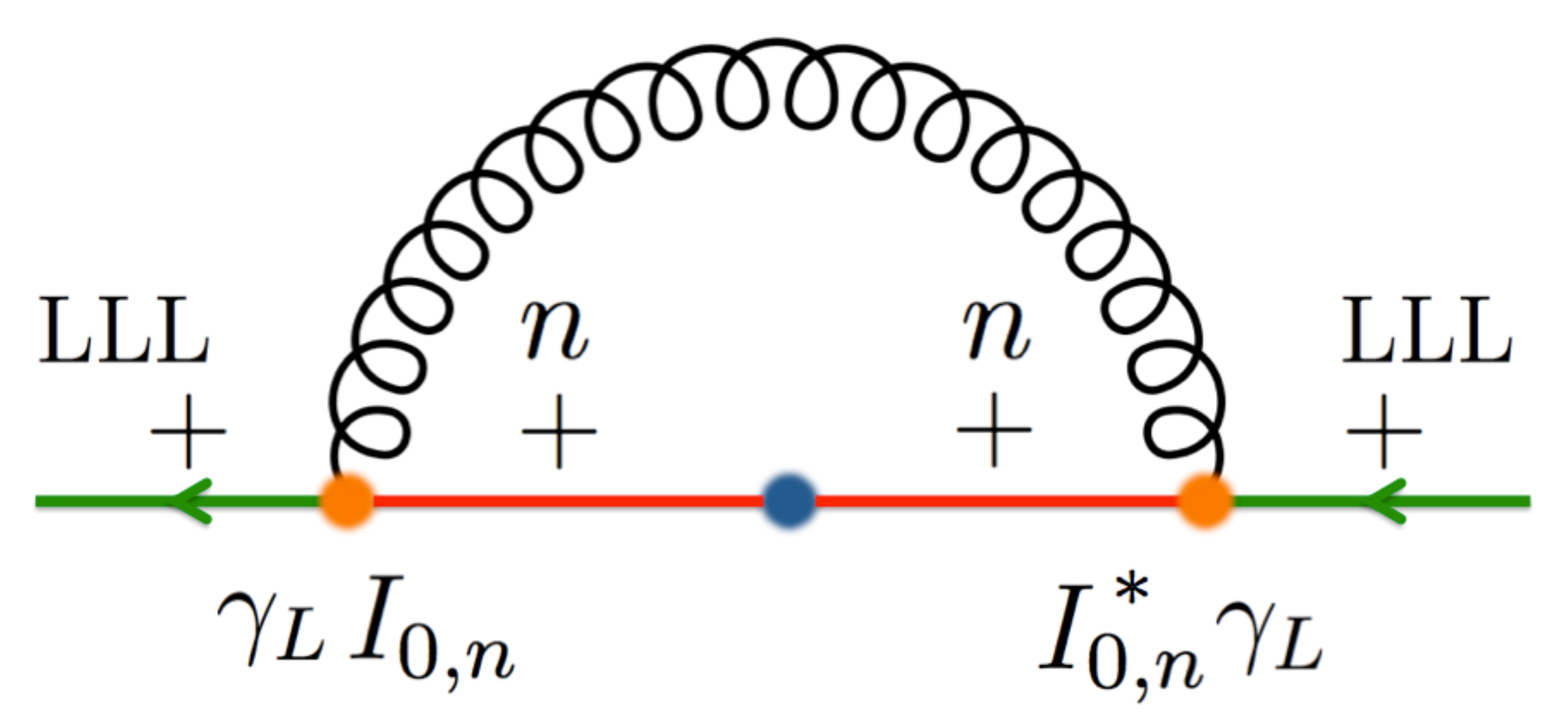} 
} \hspace{1.0cm}
\scalebox{0.6}[0.6] {
  \includegraphics[scale=.35]{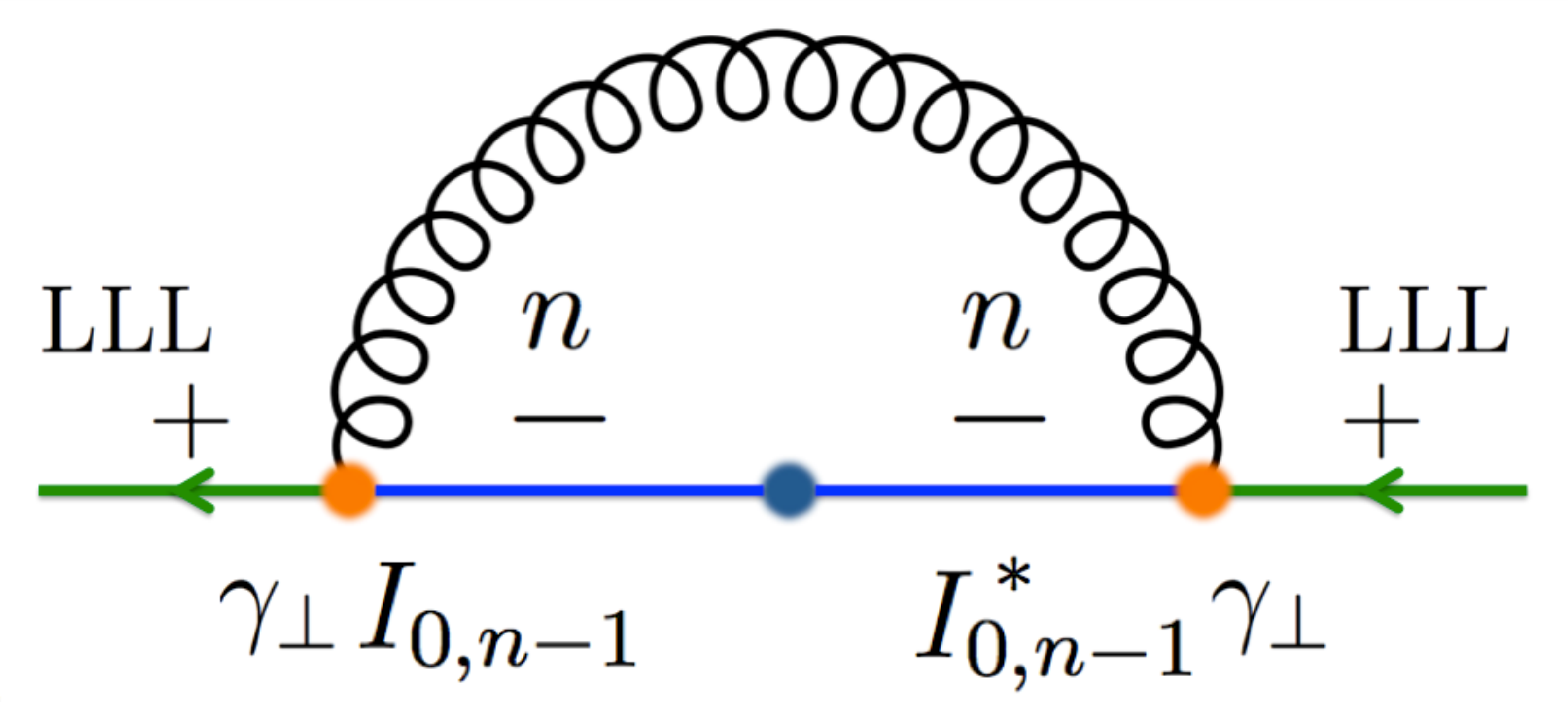} }
\end{center}
\vspace{-0.6cm}
\caption{
The one-loop corrections from the hLLs to
the LLL. 
The Feynman gauge is chosen
so that $D_{\mu \nu} \propto g_{\mu \nu}$.
The summation over $n\ge 1$ is implicit.
(Left) The spin conserving process.
(Right) The spin flipping process.
For the $n=1$ case, the soft gluon may appear
and non-perturbative effects
must be taken into account.
}
\label{fig:LLL_hLL}
\end{figure}
Before jumping to the results,
it may be useful to show some example 
of computation based on the Ritus base.
Below we use the gluon propagator in the Feynman gauge,
$D_{\mu \nu}^{ab} (k) = g_s^2 \delta^{ab} g_{\mu \nu} /\, k^2 $,
for simplicity.
In this gauge we have only two types of diagrams
shown in Fig.{\ref{fig:LLL_hLL}.
Below we demonstrate computations for 
the spin conserving process
($C_F= (\Nc^2-1)/2\Nc$):\footnote{Our convention
for the self-energy is $S^{-1}=S_0^{-1} + \Sigma$,
so that $S= S_0 + S_0(-\Sigma) S_0 +\cdots$.
}
\begin{equation}
\delta \Sigma_{B}^{ {\rm sc} } (p_L)
=C_F g_s^2 \, \sum_{n=1}
\int_{k_L}
\frac{ \gamma_{L'}\left[\,  
\rmi \left( \Slash{p}_L - \Slash{k}_L \right) + m \,
\right] \gamma_{L'} }
{\, (p_L-k_L)^2 + 2n |B| + m^2 \,}
\int_{k_\perp}
\frac{1}{ \, n! \,}
\left( \frac{ k_\perp^2 }{\, 2 |B| \,} \right)^{\! n} 
\rme^{- \frac{ k_\perp^2}{ 2|B|} } \, 
\frac{1 }{\, k^2 \, } 
\,,
\end{equation}
where (\ref{formula}) 
is used and the complex phases of form factors cancel out.
The sum starts from $l=n=1$
(For the spin flipping process, the sum starts
from $l=n-1=1$).
A few remarks are in order:
(i) Since the quark propagator is independent of
transverse momenta,
the $k_\perp$ integral could be factorized;
(ii) The form factor appears as powers of 
$(k_\perp/2|B|)^n$, so that the IR part of the
gluon propagator is suppressed.
A maximum of the form factor
appears at $k_\perp^2 = 2n|B|$;
(iii) The $\gamma_2$ matrix part of the quark 
propagator drops off due to the projection operator $\calP_+$
attached to $\chi$-fields: $\calP_+ \gamma_2 \calP_+ =0$.
Furthermore, although specific to this process,
only the term proportional to $m$ will survive
after manipulating the $\gamma$-matrices.

As usual, we use the Feynman parameter $\zeta$ 
to integrate over $k_L$, and then get
\begin{equation}
\delta \Sigma_{B}^{ {\rm sc} } (p_L)
= m\, \frac{\, C_F \alpha_s\, }{\, 2\pi\, } 
\int_0^1 \rmd \zeta ~ \delta K^B_\zeta\,,
\end{equation}
where we have changed variables as $x = 2|B| k_\perp^2$,
and defined
\begin{equation}
\delta K^B_\zeta = 
\int_0^\infty \! \rmd x\,
\sum_{n=1} \,
\frac{\, x^{n} \,}{ n! }\, \rme^{- x} \, 
\frac{\, 1 \,}
{\,   \Delta_\zeta  
+ (1-\zeta) x + \zeta n  \,} \,,
~~~~~
\Delta_\zeta 
= \frac{\, \zeta (1-\zeta) p_L^2 + \zeta m^2 \,}{ 2|B| } \,.
\end{equation}
The integral in $\delta K^B_\zeta$
contains two integration variables
$(x, n)$.
We reduce the integral from
two to one dimension via proper time representation:
\begin{align}
\delta K^B_\zeta =\,
&\int_0^\infty \!\rmd x \, \sum_{n=1} \,
\frac{\, x^{n} \,}{ n! }\, \rme^{- x} \, 
\int_0^\infty  \rmd \tau ~
\rme^{ - \tau \left[  \Delta_\zeta + (1-\zeta ) x + \zeta n \right] } \, 
\nonumber \\
=\, & 
\int_0^\infty  \rmd \tau ~
\rme^{ - \tau \Delta_\zeta  } 
\int_0^\infty \rmd x ~
\rme^{- x \left[ 1 + (1-\zeta) \tau \right] } \, 
\left( \sum_{n=0} \,
\frac{\, \left( x \, \rme^{- \tau \zeta} \right)^{n} \,}{ n! }
- 1 \right)\, 
\nonumber \\
=\, & 
\int_0^\infty  \rmd \tau ~ \rme^{ - \tau \Delta_\zeta  } 
\int_0^\infty \rmd x ~
\left(
\rme^{- x \left[ 1 + (1-\zeta) \tau - \rme^{-\tau \zeta} \right] } \, 
-
\rme^{- x \left[ 1 + (1-\zeta) \tau \right] } 
\right)
\nonumber \\
=\, & 
\int_0^\infty  \rmd \tau ~
\left(
\frac{ \rme^{ - \tau \Delta_\zeta  } }
{\,  1 - \rme^{-\tau \zeta} + (1-\zeta) \tau \, } 
- \frac{ \rme^{ - \tau \Delta_\zeta  } }
{\,  1 + (1-\zeta) \tau \, } \right)
\,.
\label{integralsc3}
\end{align}
The first term is singular at small $\tau$
which will be regulated by subtracting 
the vacuum piece.

\subsection{Phase space separation at $B=0$}

Having seen how the form factor function 
effectively restricts the integral over $k_\perp$,
it is natural to identify
$\Sigma_{ {\rm vac} }^{ {\rm sc},\, l=0}$ as
(odd ${k}_\perp$ terms are dropped)
\begin{equation}
\Sigma_{ {\rm vac} }^{ {\rm sc},\, l=0 } (p_L)
= C_F g_s^2 \, \int_{k_L} \int_{k_\perp}
\gamma_{L'}
\frac{ \rmi \left( \Slash{p}_L - \Slash{k}_L \right) 
+ m }
{\, (p_L-k_L)^2 + k_\perp^2 + m^2 \,}
\gamma_{L'} \times 
\rme^{- \frac{ k_\perp^2}{ 2|B|} } \, 
\frac{1 }{\, k^2 \, } 
\label{natural}
\,,
\end{equation}
which is responsible for the $l=0$ phase space,
as seen from the fact that 
the $k_\perp^2$ integral is effectively cutoff by $\sim 2|B|$.
We can get $\delta \Sigma_{ {\rm vac} }^{ {\rm sc} }$
by replacing\footnote{To separate more orbital levels up to 
$l=N$, we should subtract 
$\sum_{l=0}^N \rme^{-k_\perp^2/2|B|} /l! \times (k_\perp^2/2|B|)^l$.
} $\rme^{-k_\perp^2/2|B|} \rightarrow 1-\rme^{-k_\perp^2/2|B|}$.
The integral can be computed as before,
and the resulting expression is
\begin{equation}
\left(\,  \Sigma_{ {\rm vac} }^{ {\rm sc},\, l=0 } \,,
\delta \Sigma_{ {\rm vac} }^{ {\rm sc} } \, \right)
= m\, \frac{\, C_F \alpha_s\, }{\, 2\pi\, } 
\int_0^1 \rmd \zeta \int_0^\infty \rmd \tau~ 
\rme^{ - \tau \Delta_\zeta  } 
\left( \frac{1}{\,1+\tau \,}\,,\, \frac{1}{\, \tau \,}- \frac{1}{\,1+\tau \,} \right)\,.
\end{equation}
Here $\delta \Sigma_{ {\rm vac} }^{ {\rm sc} }$ is singular at small $\tau$
and becomes well-defined only after combining with
the bare self-energy $\delta \Sigma_{B}^{ {\rm sc} }$ at finite $B$.
On the other hand, $\Sigma_{ {\rm vac} }^{ {\rm sc},\, l=0 }$
blows up for small $p_L^2$, invalidating perturbation theory.
It will be combined with the renormalized self-energy at $B=0$:
\begin{equation}
\Sigma_{ {\rm vac} }^{R,\, {\rm sc} } (p_L; \mu)
= - m \, \frac{\, C_F \alpha_s\, }{\, 2\pi\, } 
\int_0^1 \rmd \zeta \, 
\ln \frac{\, (1-\zeta) \, p_L^2 + m^2 \, }
{\, (1-\zeta) \, \mu^2 + m^2 \, } 
\,.
\end{equation}
Similar to $\Sigma_{ {\rm vac} }^{ {\rm sc},\, l=0 }$,
this expression itself cannot be used
for small $p_L^2$:
The large logarithm
must be avoided by taking $\mu^2 \sim p_L^2$,
but such small $\mu$ leads to large $\alpha_s(\mu)$,
invalidating perturbation theory\footnote{As for
the scale setting problems, see \cite{Wu:2013ei}.}.
In contrast, after subtracting 
$\Sigma_{ {\rm vac} }^{ {\rm sc},\, l=0 }$ from this expression,
it is possible to make the expression for 
$\delta \Sigma_{ {\rm vac} }^R$ valid for all $p_L^2$.
To see this,
note that (for simplicity we ignore $m$ inside of the logarithm)
\begin{equation}
\Sigma_{ {\rm vac} }^{ {\rm sc},\, l=0 } 
\sim 
\int_0^{\Delta_\zeta^{-1}} \! \rmd \tau ~
\frac{\, 1\,}{\, 1+\tau \,} 
=  \ln \,\frac{\, \zeta (1-\zeta) p_L^2  + 2|B| \,}
 { \zeta (1-\zeta) p_L^2  } \,,
\end{equation}
this is combined with $\Sigma_{ {\rm vac} }^{ R,\, {\rm sc} }$ to yield
\begin{equation}
\delta \Sigma_{ {\rm vac} }^{R,\, {\rm sc} } 
\sim 
-\int_0^1 \rmd \zeta \, 
\ln\, \frac{\, \zeta (1-\zeta) p_L^2 + 2|B| \,}
  {\, \zeta (1-\zeta) \mu^2  \,} 
~ \rightarrow ~
- \ln \frac{\, 2|B| \rme^2 \,}{\, \mu^2 \,} \,
~~~(p_L^2 \rightarrow 0)\,.
\label{deltaSigma}
\end{equation}
In this expression,
the large logarithm can be suppressed
even for small $p_L^2$, by taking $\mu^2$
to be $\sim 2|B| \rme^2 \simeq 15 |B|$
(more precise numerical computations
give a smaller value, $\simeq 8.8 |B|$, but it is still large).
At the same time, if $2|B| \rme^2 \gg \lqcd^2$,
the perturbative expression for the renormalized
parameters $\alpha_s(\mu), m(\mu), \cdots,$ remains valid.
This result is rather natural because
we have effectively introduced the IR cutoff by 
separating contributions from the $l=0$ levels.
The appearance of the big factor ${\cal O}(10)$ 
has some resemblance to
finite temperature calculations
where the $\mu$ appropriate for perturbation theory 
has been supposed to be 
$\sim 2\pi T$ \cite{Blaizot:2003tw}.

The above estimate for $\mu$ is 
based on the rather crude approximation,
so we numerically search for the renormalization scale
such that
\begin{equation}
\delta \Sigma_{ {\rm vac} }^R (p_L, \mu^*) = 0 \,,
~~~~~ \mu^* = \mu^* \left(p_L,|B|\right) \,,
\end{equation}
where the logarithmic term disappears.
Clearly $\mu^*$ is a function of $p_L$ and $B$,
and we plot its behavior in Fig.\ref{fig:op_mu}.
Remarkably, even for the $p_L^2=0$ case,
$\mu^*$ reaches $1\, {\rm GeV}$ already at 
$|B|\simeq 0.12\, {\rm GeV}^2$.}
Since we shall soon show that the other contribution,
$\delta \Sigma_B - \delta \Sigma_{ {\rm vac} }$, are 
at most of $\sim C_F \alpha_s/2\pi \times {\cal O}(1)$ 
irrespective to values of $p_L^2/|B|$,
one can justify the perturbative
expressions of $\delta \Sigma_B^R$
as far as $\alpha_s(\mu^*)$ is small. 
Therefore we suppose that the contributions from higher orbitals
do not strongly affect the LLL already at $|B|$ of $(0.1-0.3)\, {\rm GeV}^2$.

\begin{figure}[tb]
\vspace{0.0cm}
\begin{center}
  \includegraphics[scale=.28]{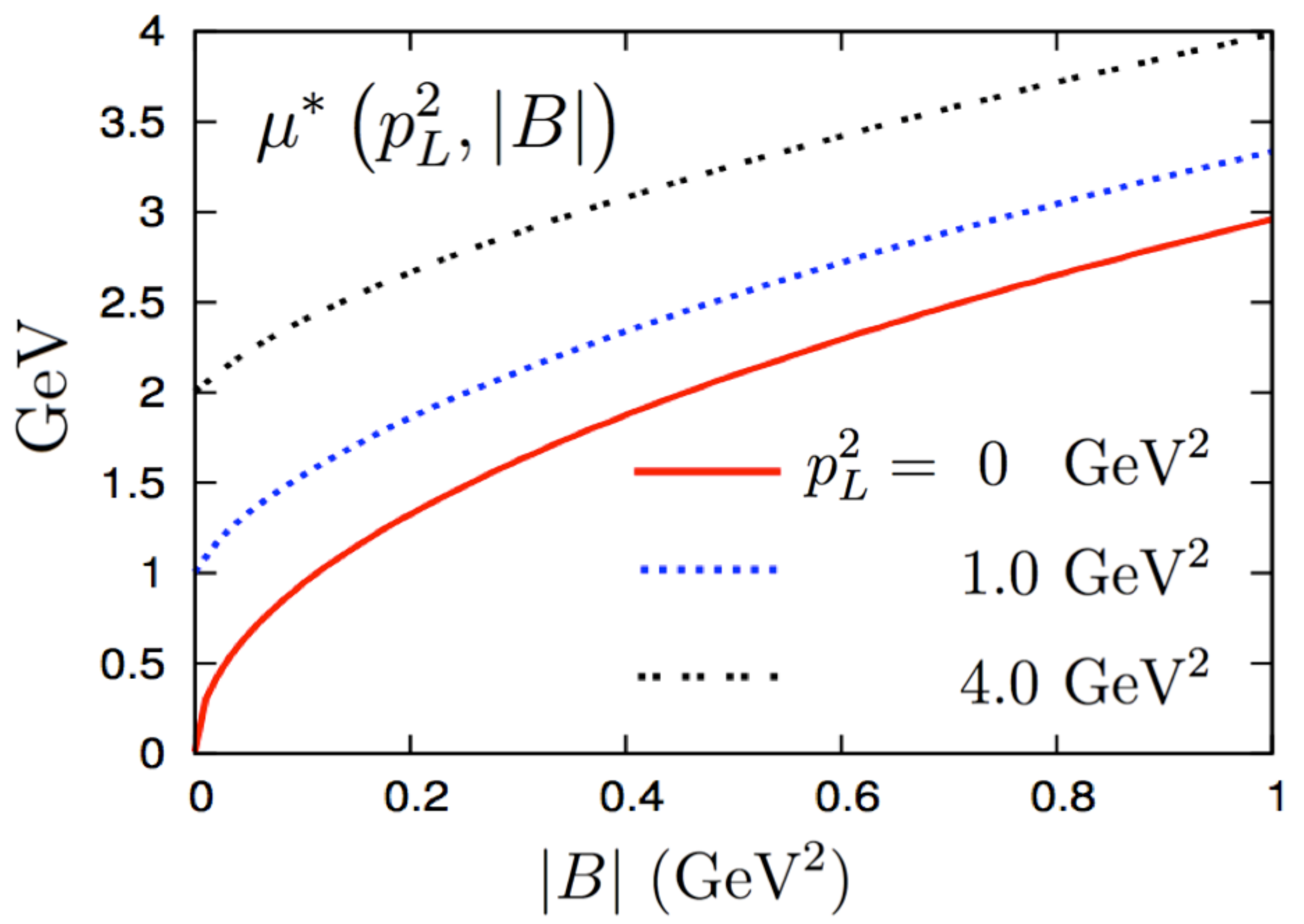} 
\end{center}
\vspace{-0.6cm}
\caption{
The optimized renormalization scale $\mu^* (p_L^2, |B|)$
which satisfies $\delta \Sigma^R_{ {\rm vac} }(p_L;\mu^*)=0$.
At $|B| \simeq 0.12\, {\rm GeV}^2$,
$\mu^*$ starts to go beyond $1.0\, {\rm GeV}$
for all $p_L^2$.
}
\label{fig:op_mu}
\end{figure}
%

\subsection{Results}
%
\begin{figure}[tb]
\vspace{0.0cm}
\begin{center}
  \includegraphics[scale=.28]{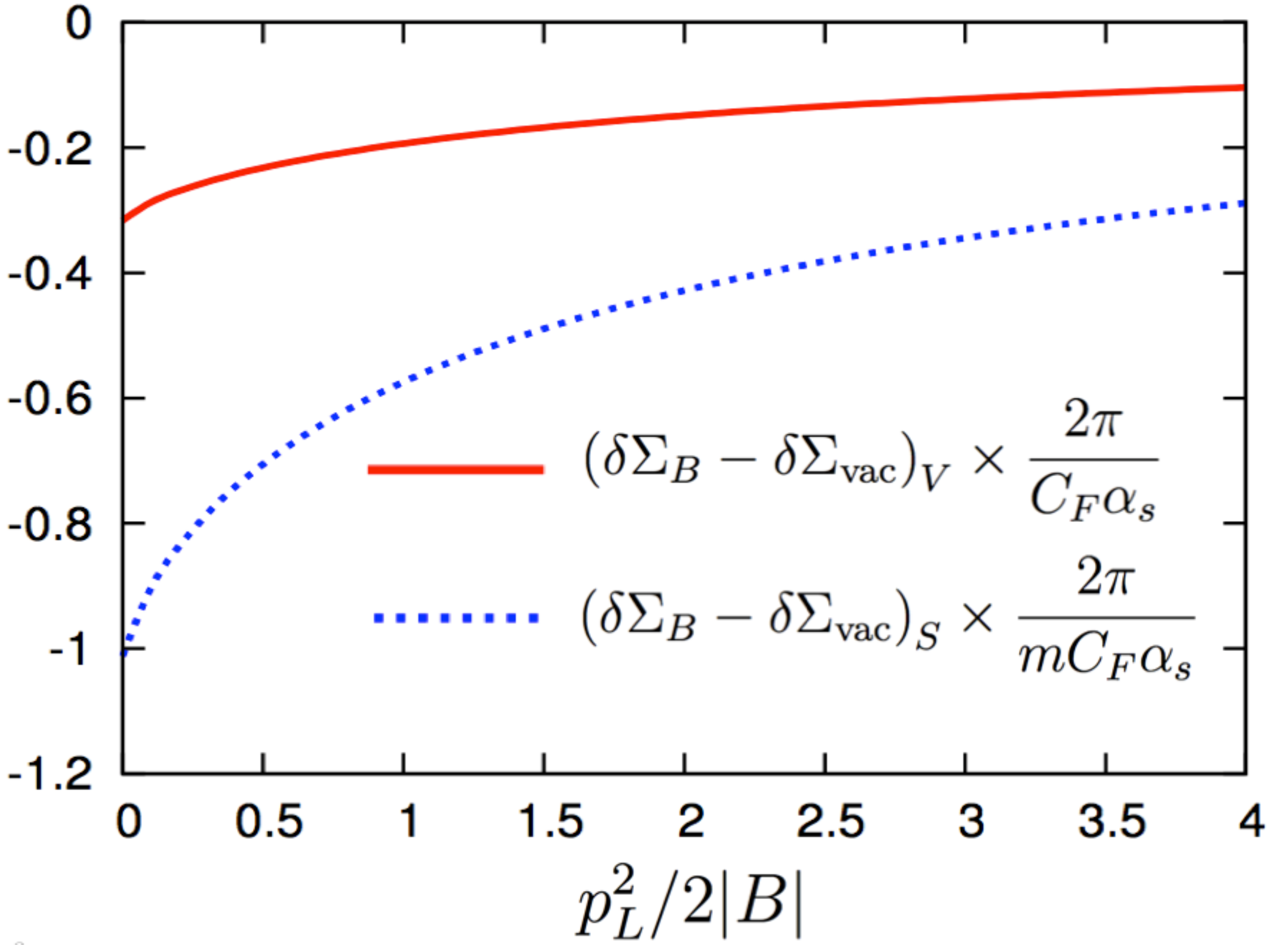} 
\end{center}
\vspace{-0.6cm}
\caption{
The $\delta \Sigma_B(p_L) - \delta \Sigma_{ {\rm vac} }(p_L)$
for vector and scalar parts.
To plot the $\mu$-independent part,
we multiply factors $2\pi/ C_F \alpha_s$ and 
$2\pi/ m C_F \alpha_s$, respectively, and then set $m=0$.
The main message is that these functions are regular and small
in the entire range of $p_L^2/2|B|$.
}
\label{fig:difsigma}
\end{figure}

Now we examine the self-energy at finite $B$
within the Feynman gauge.
We write the vector and scalar self-energies
separately,
$\Sigma = - \rmi \Slash{p}_L \Sigma_V + \Sigma_S$.
Adding contributions from the spin conserving and
flipping processes,
we arrive at expressions\footnote{In order
to make the actual numerical calculations stable,
it is more convenient to replace the log term as
\begin{equation}
-\ln \frac{\, (1-\zeta) \, p_L^2 + m^2 \, }
{\, (1-\zeta) \, \mu^2 + m^2 \, } 
= \int_0^\infty \!\rmd \tau~ 
\frac{\, \rme^{ - \tau \Delta_\zeta  } 
- \rme^{ - \tau \Delta'_\zeta  }\,}{\,\tau \,} \, 
\end{equation}
where 
$\Delta'_\zeta = [ \zeta(1-\zeta) \mu^2 + \zeta m^2 ]/2|B|$.
In this form, it is clear that the integral for
$\delta \Sigma_{ {\rm vac} }^R$
is regular in both small and large $\tau$ domains.
},
\begin{align}
\left(\delta \Sigma_{B }^R \right)_V
& = \, \frac{\, C_F \alpha_s\, }{\, 2\pi\, } 
\int_0^1 \rmd \zeta ~ (1-\zeta) \,
\bigg\{ 
\left[\,
-\ln \frac{\, (1-\zeta) \, p_L^2 + m^2 \, }
{\, (1-\zeta) \, \mu^2 + m^2 \, } 
- \int_0^\infty \!\rmd \tau~ 
\frac{\, \rme^{ - \tau \Delta_\zeta  } \,}{\,1+\tau \,} \,
\right] 
\nonumber \\
&~~~~~
+ \int_0^\infty \rmd \tau~ \rme^{ - \tau \Delta_\zeta  }
\left[\,
\frac{ \rme^{ - \tau \zeta  } }
{\,  1 - \rme^{-\tau \zeta} + (1-\zeta) \tau \, } 
- \frac{ \rme^{ - \tau \zeta  } } {\,  1 + (1-\zeta) \tau \, } 
- \frac{1}{\, \tau \,} + \frac{1}{\, 1+ \tau \,}
\right]
\bigg\}
\,,
\end{align}
\begin{align}
\left( \delta \Sigma_{B }^R \right)_S
& =  m \,\frac{\, C_F \alpha_s\, }{\, 2\pi\, } 
\int_0^1 \rmd \zeta~
\bigg\{\, 2\left[\, 
-\ln \frac{\, (1-\zeta) \, p_L^2 + m^2 \, }
{\, (1-\zeta) \, \mu^2 + m^2 \, } 
- \int_0^\infty \!\rmd \tau~ 
\frac{\, \rme^{ - \tau \Delta_\zeta  } \,}{\,1+\tau \,} \,
\right]
\nonumber \\
&~~~~~
+ \int_0^\infty \rmd \tau~ \rme^{ - \tau \Delta_\zeta  }
\left[\,
\frac{ 1+ \rme^{ - \tau \zeta  } }
{\,  1 - \rme^{-\tau \zeta} + (1-\zeta) \tau \, } 
- \frac{ 1+ \rme^{ - \tau \zeta  } } {\,  1 + (1-\zeta) \tau \, } 
- \frac{2}{\, \tau \,} + \frac{2}{\, 1+ \tau \,}
\right]
\bigg\}
\,,
\end{align}
where the first bracket comes from
$\delta \Sigma_{ {\rm vac} }^R$,
and the second from the difference between the 
bare self-energies,
$\delta \Sigma_{ B } - \delta \Sigma_{ {\rm vac} }$.

We have already seen that $\delta \Sigma_{ {\rm vac} }^R$
contains the logarithms largely related to
the choice of the renormalization scale and 
can be eliminated at $\mu^*$.
So let us look at the remaining contributions, i.e. the
$(\delta \Sigma_{ B } - \delta \Sigma_{ {\rm vac} })$'s,
which contain rather complicated integrals
over $\tau$ and $\zeta$.
We examine magnitudes and signs of the
$(\delta \Sigma_{ B } - \delta \Sigma_{ {\rm vac} })$'s
as shown in Fig.\ref{fig:difsigma}.
To plot the $\mu$-independent part,
we multiply
$2\pi/ C_F \alpha_s$ ($2\pi/ m C_F \alpha_s$)
to the vector (scalar) part and then set $m(\mu)=0$.
(Below we will always set $m=0$ 
  inside the integrals.
Introduction of the mass just makes
perturbation theory work better.)
Their values are small and regular
in the entire domain of $p_L^2/2|B|$.
Thus perturbative expressions for the
$(\delta \Sigma_{ B } - \delta \Sigma_{ {\rm vac} })$'s
are valid as far as $\alpha_s(\mu)$ is small,
as already advertised.

These self-energies can be converted
into the field strength and mass at finite $B$ through
the following relations:
\begin{equation}
\frac{1}{\, 1+\left(\delta \Sigma_{B }^R \right)_V \,}
\equiv Z_B(p_L;\mu) \,,
~~~~~
Z_B
\left[\,
1 + \frac{\,\left( \delta \Sigma_{B }^R \right)_S \,}{m } 
\,\right]
\, m(\mu)
\equiv m_B(p_L;\mu) \,.
\end{equation}
We call $m_B$ ``current quark mass'' 
of the LLL.
For the renormalized parameters at $B=0$,
we use the 1-loop expressions for
the running coupling and mass:
\begin{equation}
\alpha_s(\mu) 
= \frac{1}{\, 4\pi \beta_0 \ln (\mu^2/\lqcd^2) \,}\,,
~~~
m(\mu) = \left( \frac{ \ln (\mu_0^2/\lqcd^2) }
{\, \ln (\mu^2/\lqcd^2)\,} \right)^{ \!\frac{\gamma_{m0}}{2\beta_0} } 
\, m(\mu_0) \,,
\end{equation}
where $\beta_0 = (11\Nc - 2\Nf)/48\pi^2$
and $\gamma_{m0} = 3 C_F/ (4\pi)^2$.
In the following,
we take $(\Nc, \Nf)=(3,\, 5)$ and use
the central value of 
$\lqcd = \Lambda_{ \overline{ {\rm MS} } }^{\Nf=5} 
= (213 \pm 8)\, {\rm MeV}$ 
\cite{Beringer:1900zz}.

\begin{figure}[tb]
\vspace{0.0cm}
\begin{center}
 \hspace{-0.3cm}
\scalebox{0.58}[0.6] {
  \includegraphics[scale=.29]{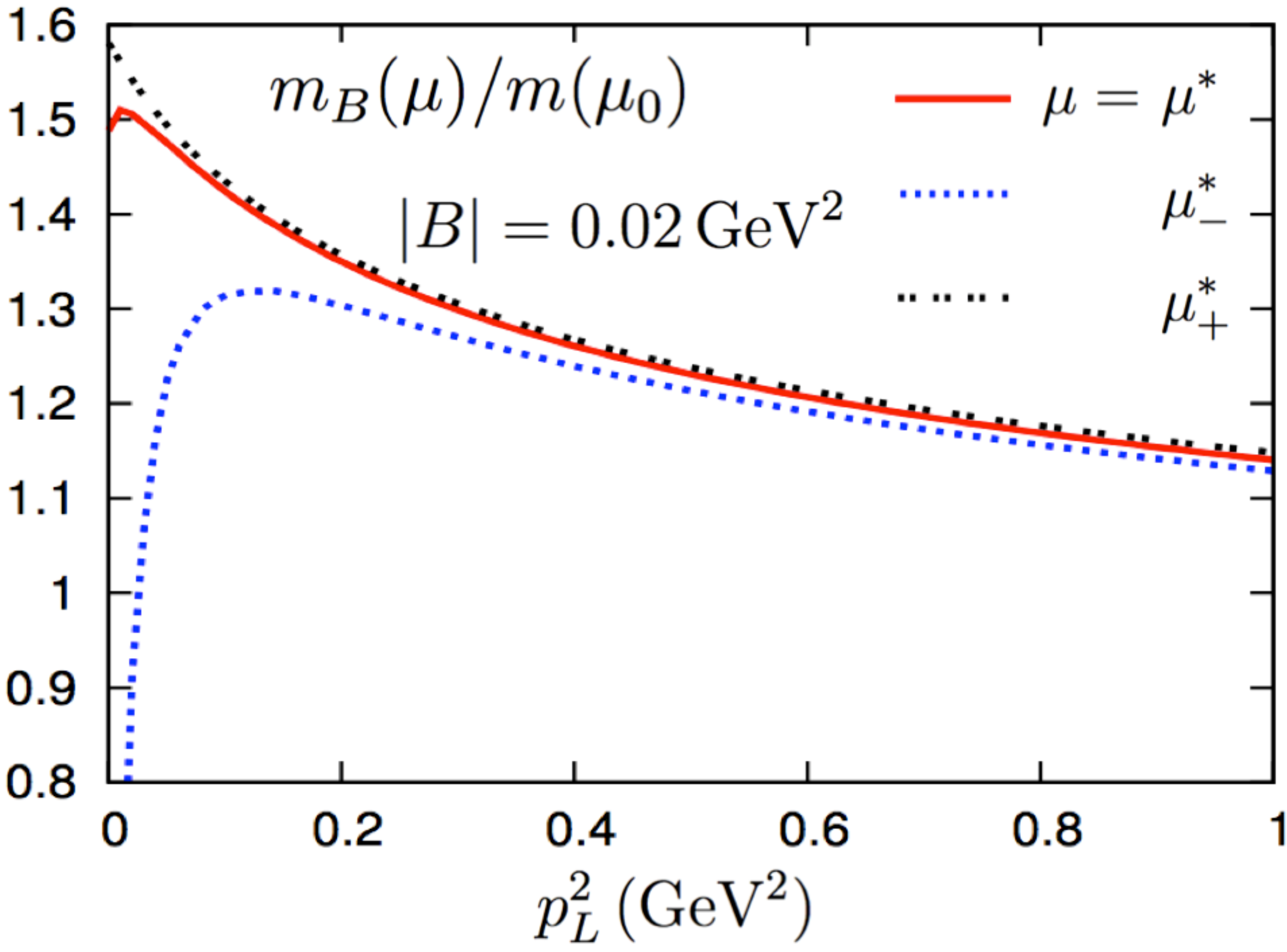} }
 \hspace{-0.3cm}
\scalebox{0.58}[0.6] {
  \includegraphics[scale=.29]{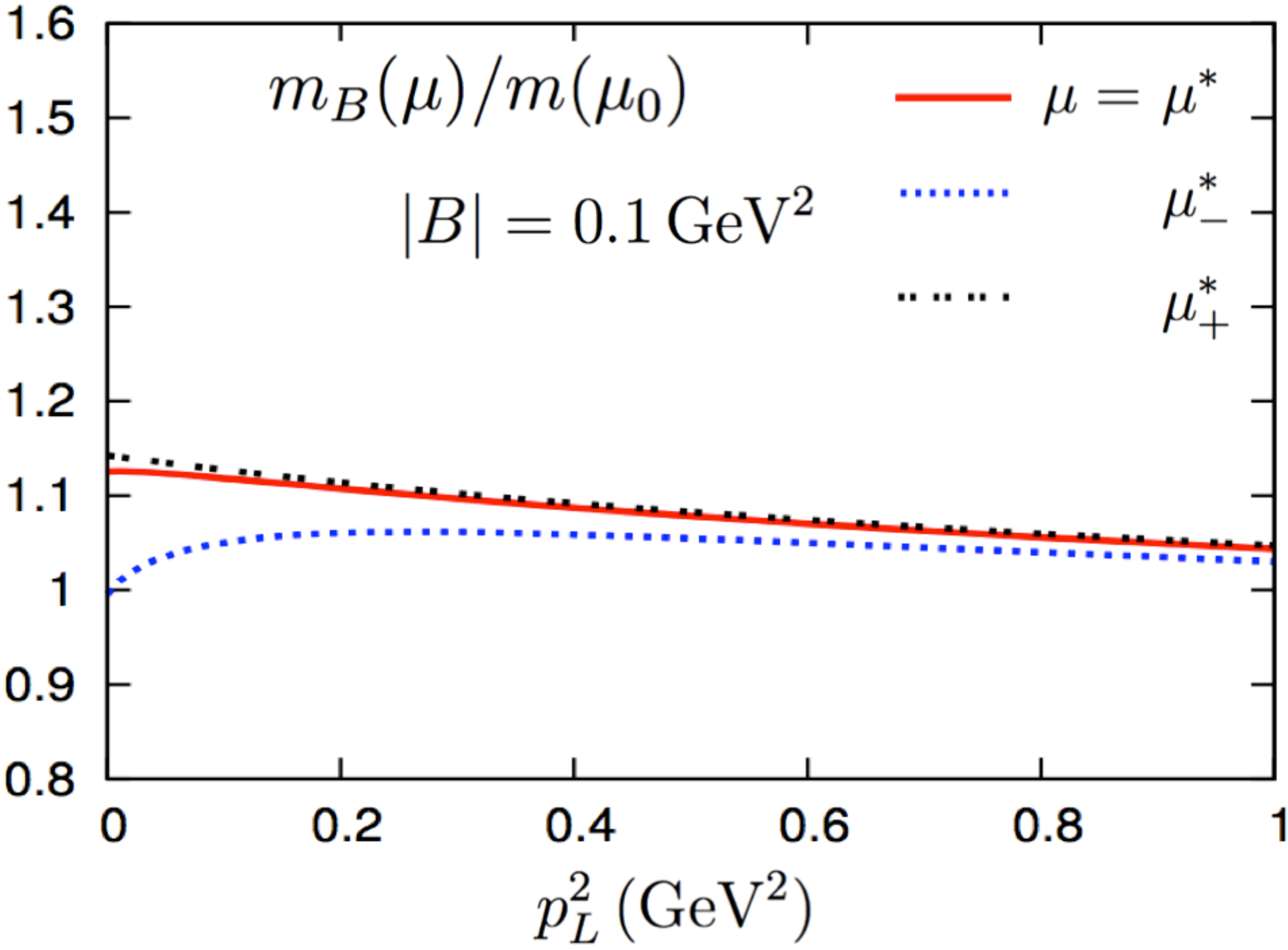} }
\hspace{-0.3cm}
\scalebox{0.58}[0.6] {
  \includegraphics[scale=.29]{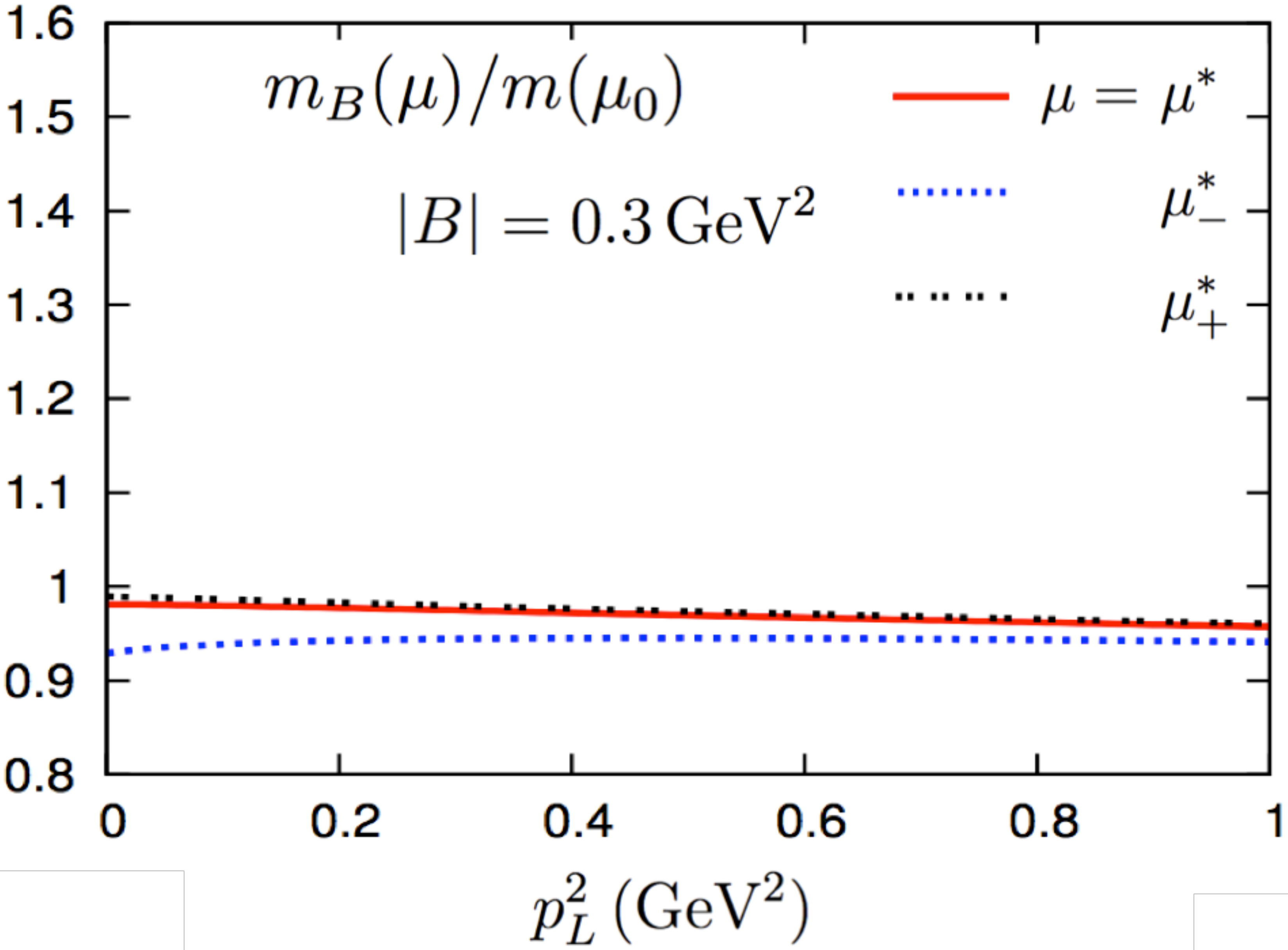} }
\end{center}
\vspace{-0.6cm}
\caption{
The ``current quark mass''  of the LLL,
$m_B$, as a function of $p_L^2$.
The function is normalized by the current quark mass
at $B=0$, $m(\mu_0)$ with $\mu_0=2\, {\rm GeV}$.
We plot the $|B| =(0.02,\, 0.1,\, 0.3)\, {\rm GeV}^2$ cases.
The renormalization scale is varied around $\mu^*$,
see discussions around Eq.(\ref{mudep}).
As $|B|$ increases,
the overall size as well as the $\mu$-dependence
become smaller.
}
\label{fig:2massB}
\end{figure}

In Fig.\ref{fig:2massB}, 
we show the behavior of $m_B$ for given $p_L^2$ and $B$.
The $B=(0.02,\, 0.1,\, 0.3)\, {\rm GeV}^2$
cases are plotted.
The function is divided by 
the current quark mass at $B=0$
renormalized at $\mu_0=2\, {\rm GeV}$.
One can get the mass functions for each flavor
by multiplying $m_f (2\, {\rm GeV})$ 
and rescaling $B_f (= e_f B)$ appropriately
for each flavor.

As for the renormalization scale,
we take $\mu$ to be $\sim \mu^*(p_L^2,|B|)$
following the discussions around Eq.(\ref{deltaSigma}).
We attach some errors by varying $\mu^*$ 
because the logarithmic terms of the higher order loops 
are not always eliminated by $\mu^*$ determined at 1-loop.
We use the following procedure:
First we suppose that $\mu^{*2}$ takes the form,
\begin{equation}
\mu^{*2}(p_L^2, |B|) = p_L^2+ C(p_L^2) |B| + \cdots,
\label{mudep}
\end{equation}
and compute $C(p_L^2)$.
Then we define
$\mu_\pm^{*2} \equiv \mu^{*2} \pm 0.5 \, C|B|$
and plot $m_B$ at $\mu_\pm^{*}$
to examine its renormalization scale dependence.
The $\mu$-dependence is rather large 
for the $B=0.02\, {\rm GeV}^2$ case
mainly due to the strong running of $\alpha_s$
at small $\mu^*$.
On the other hand,
for the $B=0.3\, {\rm GeV}^2$ case,
$\mu^*$ is already large for $p_L^2=0$, $\mu^* \ge 1\, {\rm GeV}$,
and there is no strong dependence on $\mu$.

%

\section{Conclusion}

In this Letter we have 
developed a systematic scheme for preparing
the renormalized parameters for the LLL fields from first-principle QCD,
that will be inputs for forthcoming non-perturbative analyses.
The form factors in the Ritus bases
allow us to apply perturbation theory 
in evaluating the coupling between the LLL
and higher orbital levels.
The renormalization for the quark self-energy is done
by summing up all higher orbital levels
in a model independent manner.

The results indicate that 
the higher orbital levels cease to strongly affect 
the LLL\footnote{The opposite 
is certainly not true.
Even at large $B$, the LLL can affect 
the higher orbital levels considerably.}
at rather small value of $|eB|$, $(0.1-0.3)\, {\rm GeV}^2$.
It is tempting to compare this estimate
with lattice data for the chiral condensate
where characteristic changes are observed
around $|eB| \sim 0.3\, {\rm GeV}^2$.
We conjecture that
the changes can be attributed to
the separation of the hLLs from the LLL.

To extract real phenomenological interpretations
from our perturbative results, 
however,
we still need further studies 
of other renormalized parameters\footnote{Recently,
the photon polarization at 1-loop was analyzed
in great detail, including all Landau levels \cite{Hattori:2012je}.
See also \cite{Ishikawa:2013fxa} for a related work.
The results can be applied to our framework
with a modification to separate the LLL.
}, 
effective operators, and the OPE for soft gluons.
The extension to higher loops is also a very important issue
to check whether systematics is at work.
Indeed, beyond 1-loop,
there are subdiagrams with soft gluons
even after separating the $l=0$ levels.
We leave them for future studies.

\section*{Acknowledgments}
T.K. is supported by
the Sofja Kovalevskaja program and
N.S. by the Postdoctoral Research 
Fellowship of the Alexander von Humboldt Foundation.


\end{document}